\documentclass[twocolumn,a4paper,10pt]{article}
\usepackage[T1]{fontenc}
\usepackage[latin1]{inputenc}
\usepackage[normalem]{ulem}
\usepackage{amsmath,amssymb}
\usepackage[dvips]{color}
\usepackage[dvips]{graphicx}
\usepackage{epsfig,graphicx}

\makeatletter
\renewcommand{\section}{\@startsection{section}{2}{0cm}{-\baselineskip}{0,5\baselineskip}{\normalsize\bfseries}}
\renewcommand{\subsection}{\@startsection{subsection}{3}{0cm}{-\baselineskip}{0,5\baselineskip}{\normalsize\slshape}}
\renewcommand{\subsubsection}{\@startsection{subsubsection}{3}{0cm}{-\baselineskip}{0,5\baselineskip}{\normalsize\slshape}}
\makeatother

\begin{document}

\twocolumn[
\begin{@twocolumnfalse} 
\title{\bf Production and characterization of a custom-made $^{228}$Th source with reduced neutron source strength for the Borexino experiment}

\author{W. Maneschg$^a$\footnotemark\,, L. Baudis$^b$, R. Dressler$^c$, K. Eberhardt$^d$,\\
		R. Eichler$^c$, H. Keller$^d$, R. Lackner$^a$, B. Praast$^d$, R. Santorelli$^b$,\\
        J. Schreiner$^a$, M. Tarka$^b$, B. Wiegel$^e$, A. Zimbal$^e$
\\
\\
$^a$ \small{\em Max Planck Institut f\"ur Kernphysik, Saupfercheckweg 1, D-69117 Heidelberg, Germany}\\
$^b$ \small{\em Physik Institut der Universit\"at Z\"urich, Winterthurerstrasse 190, CH-8057 Z\"urich, Switzerland}\\
$^c$ \small{\em Paul Scherrer Institut, CH-5232 Villigen, Switzerland}\\
$^d$ \small{\em Institut f\"ur Kernchemie, Universit\"at Mainz, Fritz-Strassmann-Weg 2, D-55128 Mainz, Germany}\\
$^e$ \small{\em Physikalisch-Technische Bundesanstalt, Bundesallee 100, D-38116 Braunschweig, Germany}\\
}


\maketitle

\begin{abstract}
A custom-made $^{228}$Th source of several MBq activity was produced for the Borexino experiment for studying the external background of the detector. The aim was to reduce the unwanted neutron emission produced via ($\alpha$,n) reactions in ceramics used typically for commercial $^{228}$Th sources. For this purpose a ThCl$_4$ solution was converted chemically into ThO$_2$ and embedded into a gold foil.\\ 
The paper describes the production and the characterization of the custom-made source by means of $\gamma$-activity, dose rate and neutron source strength measurements. From $\gamma$-spectroscopic measurements it was deduced that the activity transfer from the initial solution to the final source was $>$91\% (at 68\% C.L.) and the final activity was (5.41$\pm$0.30)\,MBq. The dose rate was measured by two dosimeters yielding 12.1\,mSv/h and 14.3\,mSv/h in 1\,cm distance. The neutron source strength of the 5.41\,MBq $^{228}$Th source was determined as (6.59$\pm$0.85)\,s$^{-1}$.

\end{abstract}

---------------------------------------------------------------------------------------------------------------------------------------\\
\end{@twocolumnfalse}
]

\section{Introduction}
\label{Introduction}

\renewcommand{\thefootnote}{\fnsymbol{footnote}}
\footnotetext[1]{Corresponding author.\\ {\bf e-mail}: werner.maneschg@mpi-hd.mpg.de}
\renewcommand{\thefootnote}{\arabic{footnote}}

Ceramics are often used as matrix material for the production of commercially available $\alpha$ emitters such as thorium and uranium sources. Due to the low-Z elements typically contained in ceramic materials such sources emit also  neutrons via ($\alpha$,n) reactions. The neutron source strength depends on the source activity and on the chemical composition of the ceramic.\\
Experiments in particle and astroparticle physics looking for rare events have a large interest in thorium and uranium sources for calibration measurements. However, the neutron source strengths should be reduced since the emitted neutrons can be problematic: On the one hand, they can activate detector materials and subsequent decays lead to delayed signals. On the other hand, they can cause instantaneous background signals in the own but also in neighboring experiments. Thus, a reduction of the neutron source strength is essential. This can be achieved by using materials where ($\alpha$,n) reactions are energetically not possible.\\
The present paper describes the production (Section \ref{ch:Production of a $^{228}$Th source with reduced neutron source strength}) and characterization (Section \ref{ch:Characterization of the custom-made $^{228}$Th source}) of a custom-made 5.41\,MBq $^{228}$Th source with a reduced neutron source strength for the Borexino experiment \cite{BOR08c} which is located at the Laboratori Nazionali del Gran Sasso (LNGS) in Assergi, Italy. The purpose of using such a strong source in Borexino is the determination of the external $\gamma$-background. It is caused mainly by the long-range 2.614\,MeV $\gamma$-rays originating from decays of the $^{232}$Th daughter nuclide $^{208}$Tl in the outer detector components, i.e. the stainless steel sphere, the photomultiplier tubes and the light concentrators.\\
The knowledge of the external $\gamma$-background in Borexino is of large importance for the determination of the solar neutrino rates, especially for the pep and CNO neutrino analysis but also for the measurement of the $^8$B neutrinos below the threshold of approximately 2.8\,MeV. In addition, the external calibration data are very useful for an improved understanding of the position and energy reconstruction and for comparison studies with Monte Carlo simulations in Borexino \cite{Man11}.\\

\section{Production of a $^{228}$Th source with reduced neutron source strength}
\label{ch:Production of a $^{228}$Th source with reduced neutron source strength}

\begin{figure}
\begin{center}
\includegraphics[scale=.23]{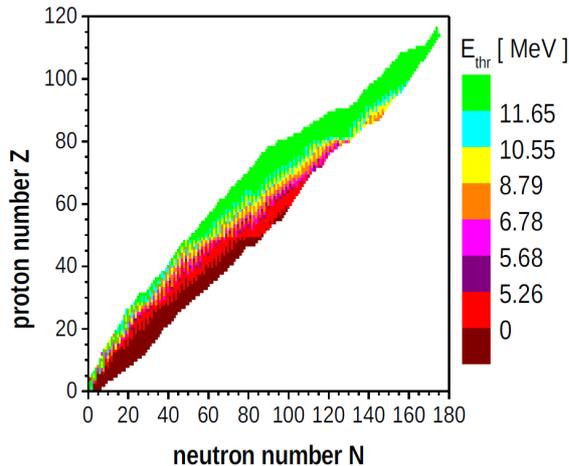}
\caption{($\alpha$,n) energy thresholds for all nuclides taken from the AME2003 atomic mass evaluation \cite{Aud03}.}
\label{fig:alpha_n}
\end{center}
\end{figure}

The technique applied for the production of the Borexino $^{228}$Th source with a reduced neutron source strength was originally developed at the Paul Scherrer Institut (PSI) in Villigen, Switzerland, in collaboration with the Physik-Institut at University of Z\"urich \cite{Tar11}, Switzerland. The method is aimed at embedding $^{228}$Th in a metallic, easily treatable material with an ($\alpha$,n) energy threshold lying above the maximal energy of $\alpha$-particles emitted from the $^{228}$Th daughter nuclides that is given by $^{212}$Po $\alpha$-particles reaching 8.8\,MeV.\\ 
Figure \ref{fig:alpha_n} shows the ($\alpha$,n) energy threshold $E{_{th}}$ for all isotopes known in 2003. For Z$<$50 only 17 stable isotopes exhibit an ($\alpha$,n) threshold $E{_{th}}$ above 8.8\,MeV. 

For the Borexino $^{228}$Th source gold was selected as a monoisotopic element which has $E{_{th}}$=9.94\,MeV.\\
For the production of the custom-made $^{228}$Th source a commercially available ThCl$_4$ solution was acquired. Its characteristics are discussed in Section \ref{ch:Characteristics of the initial ThCl$_4$ solution}. The solution underwent a chemical-thermal treatment in order to obtain ThO$_2$ which was wrapped in gold foil (Section \ref{ch:Conversion of ThCl$_4$ into ThO$_2$ on gold}). The most abundant oxygen isotope is $^{16}$O (natural abundance: 99.76\%) and is not critical since it has a threshold  $E{_{th}}$=15.17\,MeV. The two isotopes $^{17}$O and $^{18}$O have an ($\alpha$,n) energy threshold $E_{th}$ of 0.72\,MeV and 0.85\,MeV. Even though their natural abundance is low, i.e. 0.04\% and 0.2\%, the presence of these two isotopes might induce a small amount of ($\alpha$,n) reactions. Finally, the gold covered source was sealed in a double encapsulation made of stainless steel (Section \ref{ch:Encapsulation and sealing}).

\subsection{Characteristics of the initial ThCl$_4$ solution}
\label{ch:Characteristics of the initial ThCl$_4$ solution}
  
5.77\,MBq $^{228}$Th was provided by Eckert \& Ziegler Isotope Products Inc., Valencia (CA), USA, in form of thorium tetrachloride (ThCl$_4$) dissolved in 1\,M hydrochloric acid (HCl).\\
Table \ref{tab:solution-spec} summarizes the characteristics of the liquid source.\\
The amount of the synthetic radioisotope $^{229}$Th is similar to $^{228}$Th but its contribution to the total activity is negligible ($<$0.03\%). No $^{233}$U, $^{234}$U, $^{235}$U, $^{236}$U or $^{238}$U could be detected ($<$10$^{-4}$\,$\mu$g). The concentrations of non-radioactive impurities such as iron (Fe) were not determined \cite{Lev10}.\\
The solution contains also 10\,$\mu$g zirconium (Zr) which was required as precipitation carrier. Even though the carrier ions could be removed e.g. by separating the thorium tracer and zirconium carrier in 9-12\,M HCl using an anion exchange resin \cite{Hyd60}, it was not considered within the production of the Borexino source for two reasons: (1) to guarantee maximal transfer of activity from the initial to the final product; (2) the stable Zr isotopes $^{90}$Zr, $^{91}$Zr, $^{92}$Zr and $^{94}$Zr (natural abundances: 51.45\%, 11.22\%, 17.15\%, 17.38\%) have an energy threshold $E_{th}$ of 7.95\,MeV, 5.35\,MeV, 6.67\,MeV and 5.60\,MeV for ($\alpha$,n) reactions, respectively. Hence, also a small but not negligible contribution from Zr to the neutron source strength of the final product is expected.
  
\begin{table*}
\begin{center}
	\begin{tabular}{|c|cccc|}
	\hline
	Activity 				&5.77\,MBq $\pm$ 15\%		&			& Reference date:			& March 1, 2010			\\
	\hline
	Chemical composition 	&ThCl$_4$ + 1\,M HCl in 1\,ml	&		& 	  						&						\\								
	Carrier					&10\,$\mu$g Zr				&			& 	  						&						\\		
	\hline	
	Isotope composition		&Nuclide					&Weight 	& Weight by mass [\%]			&	Activity [\%]	\\
							&$^{228}$Th					&2.68\,$\mu$g	& 34.844					&	99.9703 	\\
							&$^{229}$Th					&2.92\,$\mu$g	& 39.620					&	0.0295	\\
							&$^{230}$Th					&1.86\,$\mu$g	& 25.238					&	0.0002	\\
							&$^{232}$Th					&0.022\,$\mu$g	& 0.299						&	0.0000	\\
	\hline						
	\end{tabular}
	\caption{\rm{Data sheet of the ThCl$_4$ solution delivered by Eckert \& Ziegler Isotope Products Inc., Valencia (CA), USA. The isotope composition was determined via Inductively Coupled Plasma Mass Spectrometry by the Oak Ridge National Laboratory in Oak Ridge (TN), USA.}}
	\label{tab:solution-spec}	
\end{center}
\end{table*}

\subsection{Conversion of ThCl$_4$ into ThO$_2$ on gold}
\label{ch:Conversion of ThCl$_4$ into ThO$_2$ on gold}

\subsubsection{Experimental set-up}
\label{ch:Experimental set-up}

The custom-made $^{228}$Th source for Borexino was produced at Institut f{\"u}r Kernchemie in Mainz, Germany where it is permitted to handle open radioactive material of high activity. Moreover, it provides an air filter system for collection and controlled release of radioactive gases such as $^{220}$Rn which is expected to escape permanently during the hand\-ling of the $^{228}$Th.\\
The glove box placed in a standard hood with exhaust filter contained:
\begin{itemize}
\item Teflon beaker with conical cavity and a glass fiber filter lid,
\item Precast gold crucible and a gold envelop made out of 25\,$\mu$m thick gold foil of approximately 6\,cm$^2$ area,
\item Heater covered with aluminium foil, grounded and connected to a temperature sensor located outside the glovebox,
\item Custom-made oven with temperature sensor,
\item 5\,ml and 100\,$\mu$l pipettes with polypropylene tips of type Tipac C20/TJ from GILSON,
\item HNO$_3$ acid and distilled water,
\item Real time dose meter placed outside the glovebox for monitoring the activity through the plexiglass wall of the glovebox.
\end{itemize}

Before the $^{228}$Th source for the Borexino experiment was produced the technique has been already applied for the production of three custom-made $\sim$20 kBq $^{228}$Th sources for the GERDA experiment \cite{Tar11}. Even though the technique is not complicated it is highly sensitive to environmental conditions: The transfer of activity in its final product was close to 100\% in two cases and only $\sim$35\% in one case. In order to avoid an uncontrolled loss of activity (a) an electrostatic charge of the glovebox environment and (b) larger residuals of ThCl$_4$ solution on used instrumentation have to be avoided.

\subsubsection{Procedure}
\label{ch:Procedure}

\begin{figure}
\centering
\begin{tabular}{cc}
\epsfig{file=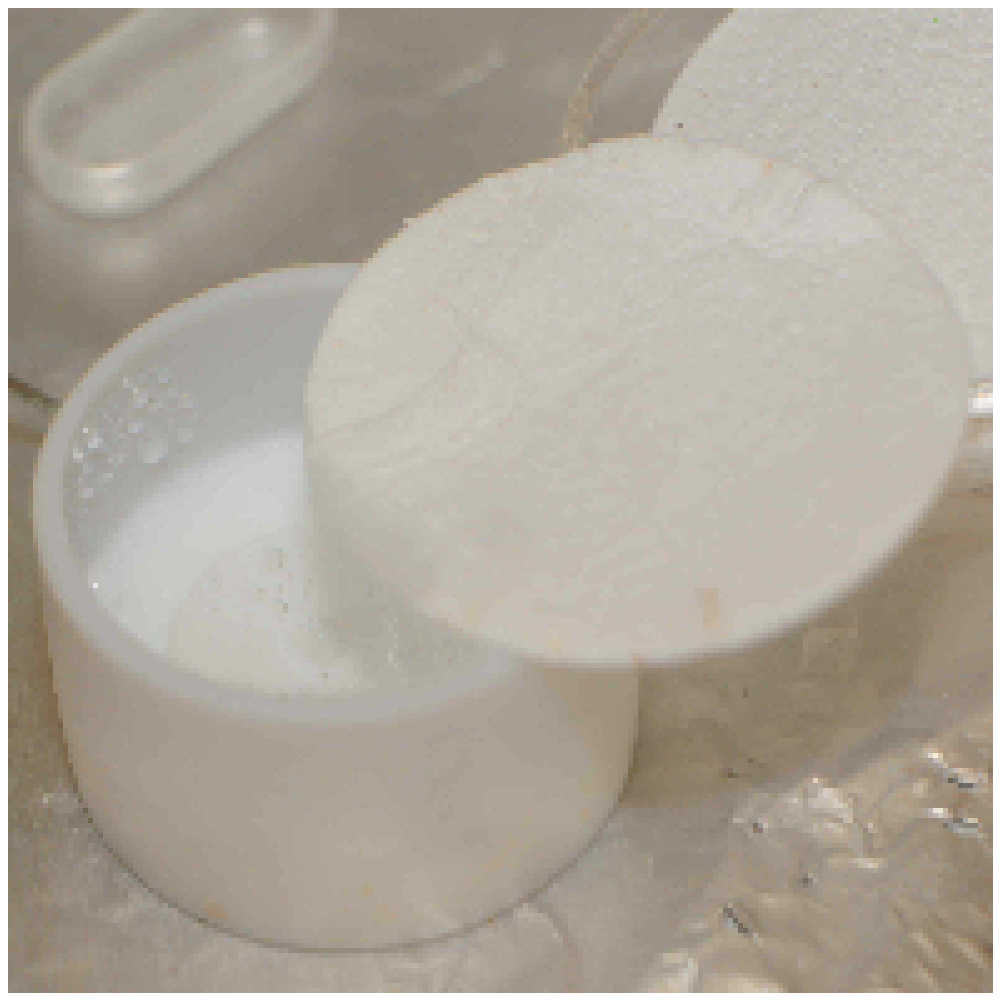,width=0.45\linewidth,clip=} & \epsfig{file=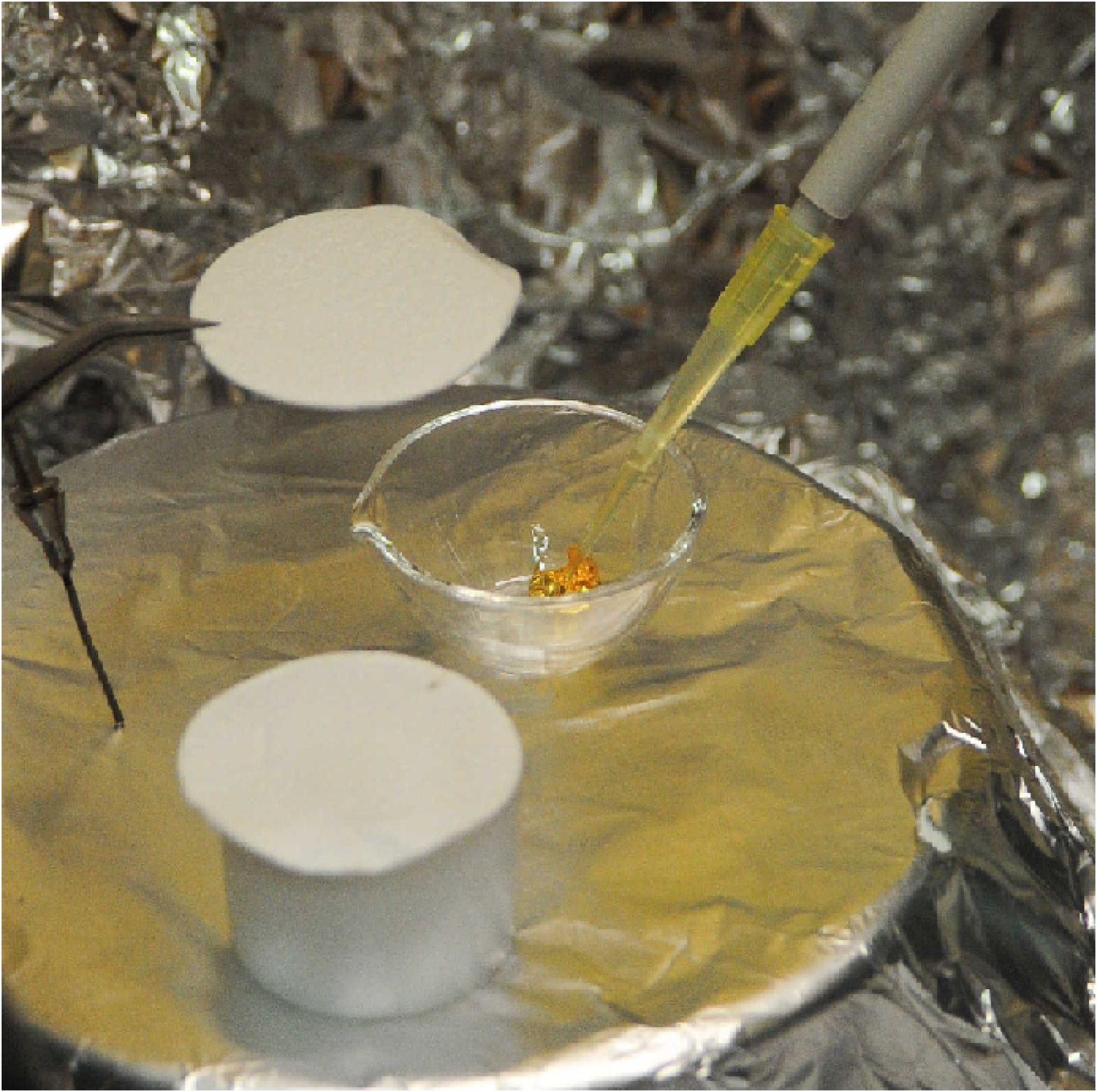,width=0.45\linewidth,clip=} \\
\epsfig{file=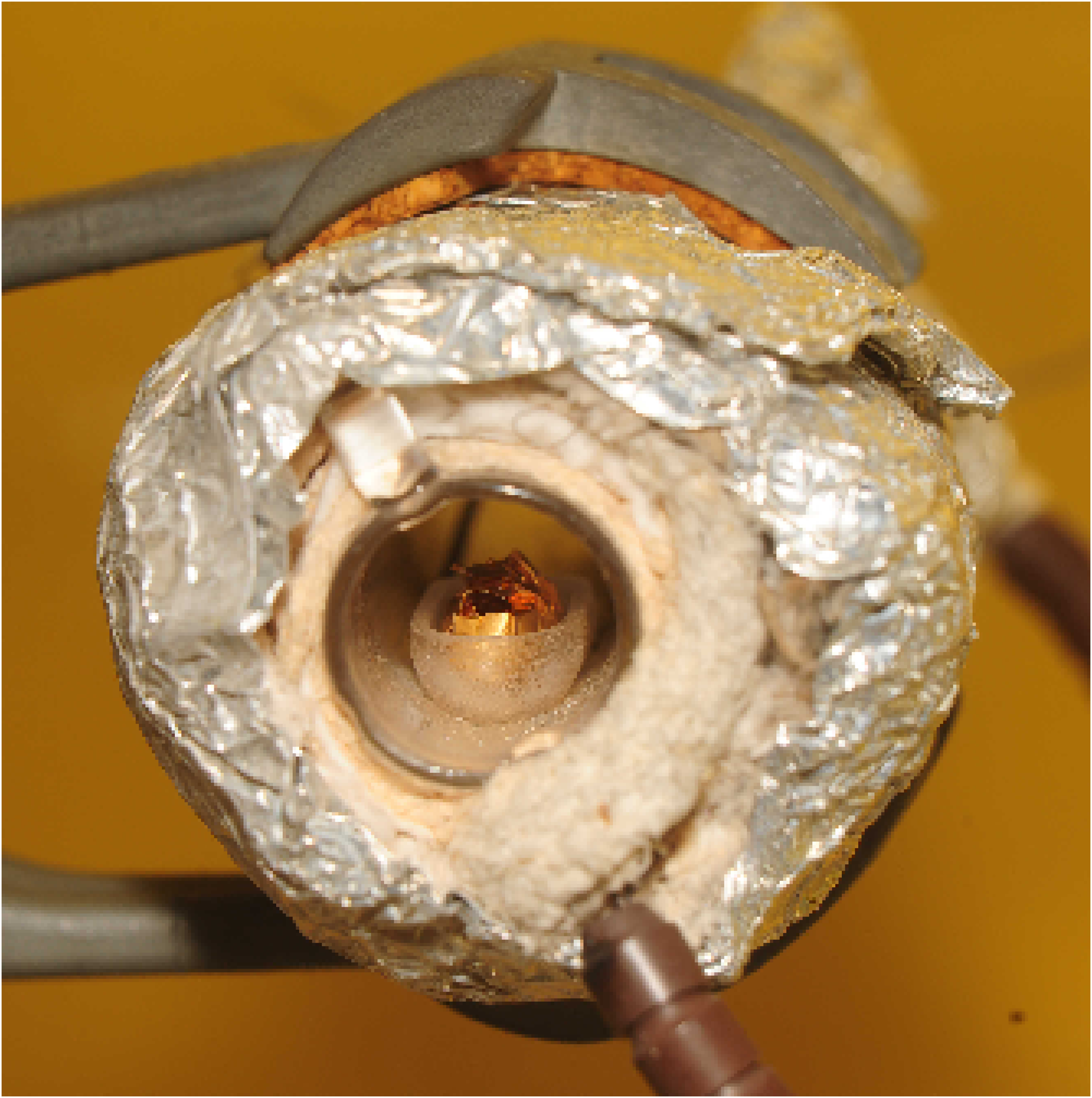,width=0.45\linewidth,clip=} & \epsfig{file=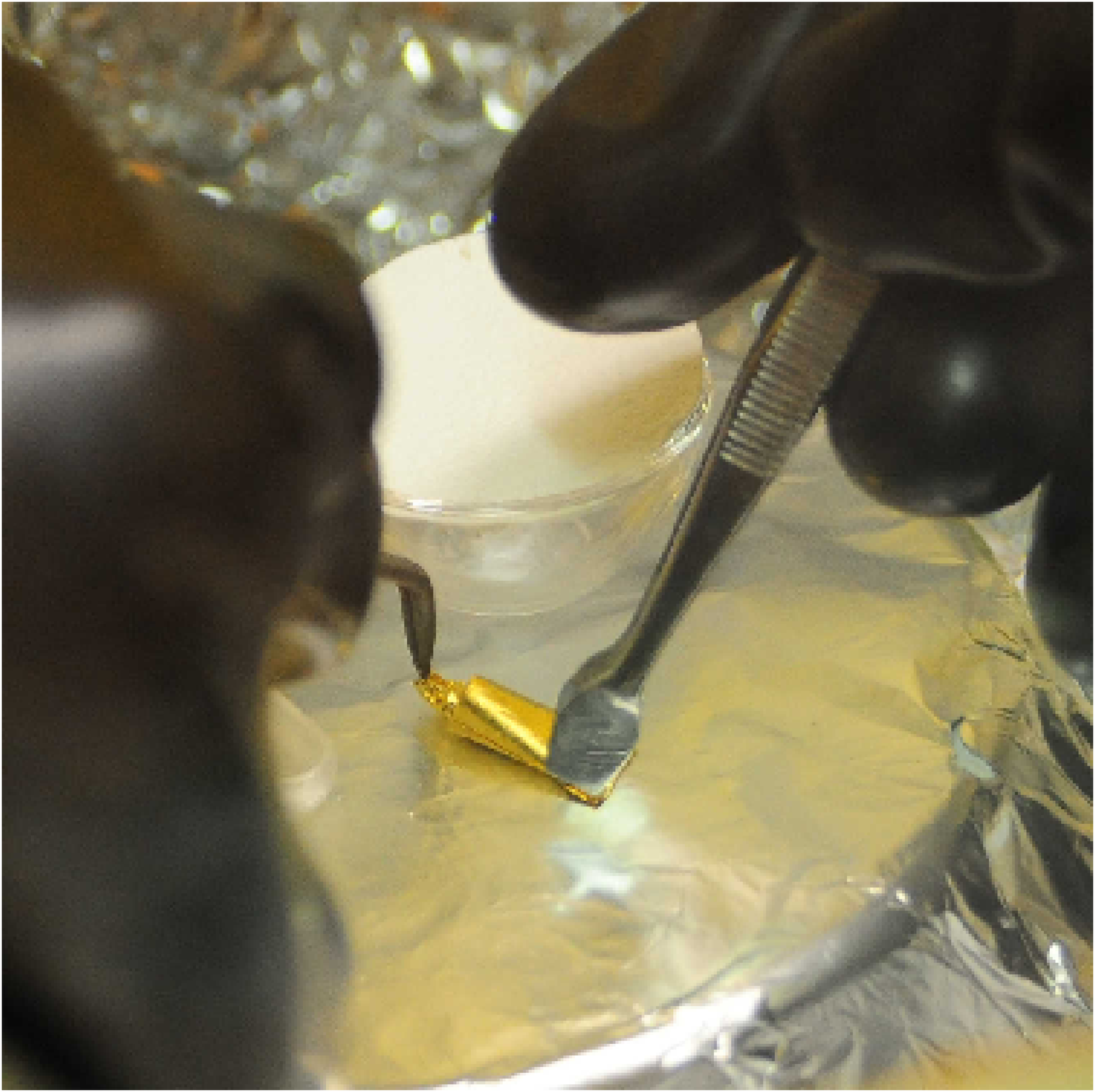,width=0.45\linewidth,clip=}
\end{tabular}
\caption{Production of the Borexino source (from left top to right bottom): Conversion of ThCl$_4$ into Th(NO$_3$)$_4$; Transfer of Th(NO$_3$)$_4$ into the gold crucible; Conversion of Th(NO$_3$)$_4$ into ThO$_2$; Insertion of the gold crucible into the gold cover.}
\label{fig:production-steps}
\end{figure}

The procedure of converting the initial ThCl$_4$ solution into ThO$_2$ in gold is as follows:
\begin{enumerate}

\item Conversion of ThCl$_4$ into Th(NO$_3$)$_4$ (Fig. \ref{fig:production-steps}, top left):\\
The ThCl$_4$ solution is transferred into a teflon beaker with conical cavity, a few ml of HNO$_3$ are added and 
evaporated to dryness at T$\sim$115$^\circ$C within $\sim$3 hours. The following reaction takes place:
  \begin{equation}  
  \begin{split}
  3\, \mathrm{ThCl_4} + 16\, \mathrm{HNO_3} \ \xrightarrow{\sim 115^{\circ}C} \\
  [3\, \mathrm{Th(NO_3)_4]_s} + [4\, \mathrm{NOCl} + 8\, \mathrm{H_2O} + 4\, \mathrm{Cl_2]_g }
  \end{split}   
  \label{eq:th-conversion-A}  
  \end{equation}
where `s' stands for solid phase and `g' for gaseous phase. Th(NO$_3$)$_4$ is generated and left in a solid state in the teflon cavity, while the remnants nitrosylchloride, water and chlorine evaporate.

\item Transfer of Th(NO$_3$)$_4$ into the gold crucible (Fig. \ref{fig:production-steps}, top right):\\
The remaining Th(NO$_3$)$_4$ was dissolved in about 100\,$\mu$l concentrated HNO$_3$ and was stepwise transferred into the small gold crucible. The crucible was heated up to T$\sim$120$^\circ$C for $\sim$30\,min to evaporate the HNO$_3$.\\
In principle, the conversion of ThCl$_4$ into Th(NO$_3$)$_4$ could be directly performed in the gold crucible. However, the handling of the small volumina in the small crucible turned out to be quite challenging due to capillarity forces. Moreover, during evaporation HNO$_3$ in combination with HCl gives a highly corrosive {\it{aqua regia}} which dissolves gold. Therefore, to prevent a spill either the gold foil must be thick enough or the process has to be bypassed.

\item Conversion of Th(NO$_3$)$_4$ into ThO$_2$ (Fig. \ref{fig:production-steps}, bottom left):\\ 
The gold crucible is folded geometrically, but not hermetically. It is inserted into an oven and heated at T$\sim$650$^\circ$C for $\sim$30\,min. At this temperature residual organic compounds volatilize and the Th(NO$_3$)$_4$ compounds are destroyed. Th(NO$_3$)$_4$ releases NO$_2$ and O$_2$ and forms ThO$_2$ on the inner surface of the gold foil according to:
  \begin{equation}
  \begin{split} 
  3\, \mathrm{Th(NO_3)_4} \ \xrightarrow{\sim 650^{\circ}C} \\
  [3\, \mathrm{ThO_2]_s} + [12\, \mathrm{NO_2} + 3\, \mathrm{O_2]_g}
  \end{split} 
  \label{eq:th-conversion-A} 
  \end{equation}
\item Folding of gold crucible (Fig. \ref{fig:production-steps}, bottom right):\\
Finally, the gold crucible is inserted in an additional gold cover and is folded several times. This prevent a loss of activity and seals contaminations on the outside of the primary gold containment.
\end{enumerate}

\subsection{Encapsulation and sealing}
\label{ch:Encapsulation and sealing}

\begin{figure}
\begin{center}
\includegraphics[scale=.20]{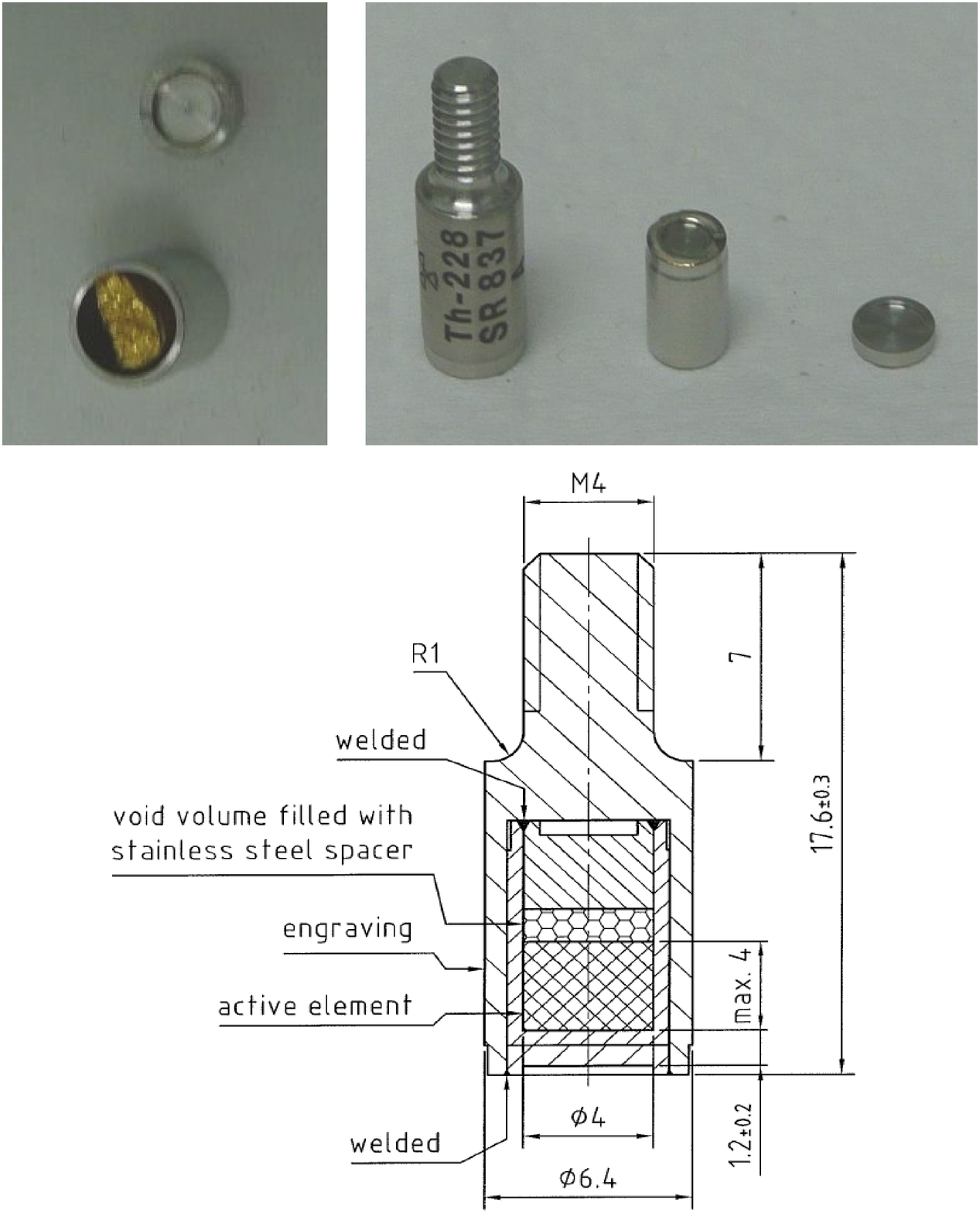}
\caption{Photos of the inclusion of the radioactive gold foil into the double encapsulation of the Borexino $^{228}$Th source. The sizes of the profile of the encapsulation are given in millimeters. Images courtesy of Eckert und Ziegler Nuclitec GmbH.}
\label{fig:encapsulation}
\end{center}
\end{figure}

The radioactive gold foil was encapsulated and sealed by the company Eckert und Ziegler Nuclitec GmbH in Braunschweig, Germany. The profile and a photo of the used double encapsulation of type VZ-3474-002 are shown in Figure \ref{fig:encapsulation}. The encapsulation has a small diameter of 6.4\,mm in order to fit into small cavities such as the re-entrant tubes of the Borexino external calibration system. In addition, one ending of the capsule has a thread which allows to fix the source on an insertion road.\\
Subsequently, leakage and contamination tests were performed. The required criteria were fulfilled, i.e. the removed activity (a) with a wipe moistened with ethanol is $<$200\,Bq and (b) after immersion of the source in a suitable liquid at 50$^\circ$C for at least 4\,h is $<$200 Bq.\\ 
An additional wipe test prior the usage of the $^{228}$Th source in the Borexino experiment at LNGS demonstrated that the upper limit of the activity on the outer surface of the source encapsulation is $<$10\,mBq.

\section{Characterization of the custom-made $^{228}$Th source}
  \label{ch:Characterization of the custom-made $^{228}$Th source}

\subsection{Gamma-spectroscopic measurements}
\label{Gamma-spectroscopic measurements}

\subsubsection{Activity transfer from the initial solution to the final product}
\label{Activity transfer from the initial solution to the final product}

A first series of $\gamma$-spectroscopic measurements were carried out at Institut f{\"u}r Kernchemie in Mainz, Germany, directly before and after the source production. The aim was to define the relative loss of activity on the instrumentation within the glovebox due to non-ideal environmental conditions (see Section \ref{ch:Experimental set-up}). In case of larger residues, it would have been mandatory to search for the radioactive residues and to reinsert them in the final source to guarantee the expected source strength for the ca\-li\-bra\-tion measurement in Borexino.\\
For the measurements an unshielded High Purity Germanium Coaxial Detector manufactured by Ortec was used. The energy range was set to [40,2000]\,keV. Its {\it full width half maximum} is 1.8\,keV at the 1.33\,MeV $\gamma$-line from $^{60}$Co.\\
The main conclusion is that the yield of activity transfer to the final $^{228}$Th source is larger than 91\% (at 68\% C.L.). Even though the result still had a large uncertainty, this interim result was satisfactory, so the custom-made source could be sent to Eckert \& Ziegler Nuclitec GmbH for encapsulation (see Section \ref{ch:Encapsulation and sealing}).

\subsubsection{Estimation of the absolute gamma activity of the source}
\label{Estimation of the absolute activity of the source}

\begin{table*}[t]
\begin{center}
	\begin{tabular}{|c|cc|c|}
	\hline
    		Nuclide		&Energy 		& Branching ratio 		&Corrected count rate 	\\
    					& [keV]			& [\%]					& [10$^6$ counts\,s$^{-1}$] \\    		
    \hline
    		$^{212}$Pb	&238.6 			&43.3					&2.05 $\pm$ 21\% 	\\ 
      		$^{208}$Tl	&583.2 			&30.4					&4.63 $\pm$ 11\% 	\\ 		
      					&2614.5 		&35.6					&5.41 $\pm$ 6\% 	\\ 
    \hline						
    \end{tabular}			
	\caption{\rm{Count rates of three prominent $\gamma$-lines from the custom-made $^{228}$Th source. The count rates are corrected for the branching ratios of the corresponding $\gamma$-lines. The corrected count rates of lower energetic $\gamma$-rays are suppressed due to self-absorption effects in the source encapsulation.}}
	\label{tab:absolute-activity}	
\end{center}
\end{table*}

\begin{figure}
\begin{center}
\includegraphics[scale=0.33,angle=270]{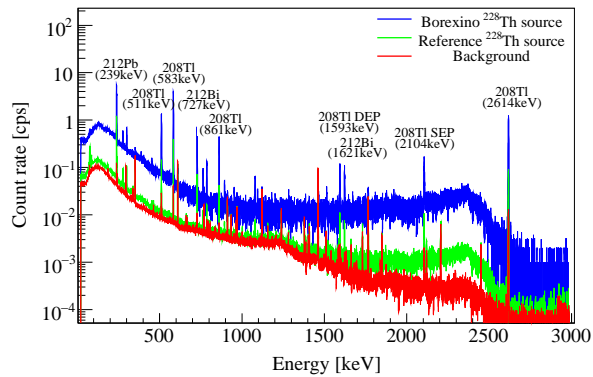}
\caption{Energy spectrum of the Borexino $^{228}$Th source and the $^{228}$Th reference source. The $\gamma$-rays contributing to the background stem from the environment and intrinsic contamination of the detector. The count rates are expressed in counts per second [cps].}
\label{fig:gamma-spectra}
\end{center}
\end{figure}

The absolute $\gamma$-activity of the Borexino $^{228}$Th source was estimated by a comparison with a ca\-li\-bra\-ted reference $^{228}$Th source by means of $\gamma$-ray spectroscopy.\\
The activity of the reference $^{228}$Th source was selected in order to exceed the background of the environment also at a larger distance from the detector, such that both $^{228}$Th sources could be treated as point-like sources. For this purpose a calibrated $^{228}$Th source with an activity of $A_{\mathrm{r}}$=297\,kBq $\pm$ 5\% (at 2$\sigma$; reference date: March 1, 2010) was selected.\\
The used $\gamma$-spectrometer was a Broad Energy Germanium detector (BEGe) manufactured by Canberra. The energy range of this detector was set to approximately [20,3000]\,keV and its {\it full width half maximum} is 1.7\,keV at the 1.33\,MeV $\gamma$-line from $^{60}$Co and 0.57\,keV at the 59.5\,keV $\gamma$-line from $^{241}$Am.\\
For the measurement the $^{228}$Th sources were placed one by one in $\sim$162\,cm distance from the unshielded detector. At this distance both sources can be assumed to be point-like. One measurement for each of the two $^{228}$Th sources and one of the background were carried out. The spectra are shown in Figure \ref{fig:gamma-spectra}.\\
The evaluation was carried out according to \cite{DIN254825}. The obtained results are summarized in Table \ref{tab:absolute-activity}.\\
The measured count rates of single $\gamma$-lines corrected by their branching ratios are not equal due to self-absorption in the source encapsulation. They differ by more than factor of 2 in the range of [238,2615]\,keV. For the high-energetic 2.615\,MeV $\gamma$-rays a corrected count rate of 
\begin{equation*}
(5.41 \pm 0.30)\times10^6 \mathrm{counts\,s^{-1}}
  \label{eq:source-gamma-activity}
\end{equation*}
is obtained (reference date: March 1, 2010). Taking into account a small activity loss during the source production and the intrinsic attenuation due to the $~$1.2\,mm thick encapsulation (few \% for 2.615\,MeV $\gamma$-rays) there is a very good agreement between the final source activity and the initial activity of the ThCl$_4$ solution (compare with Section \ref{Activity transfer from the initial solution to the final product} and Table \ref{tab:solution-spec}).

\subsection{Neutron source strength measurement} 
\label{Neutron source strength measurement} 

\begin{figure}
\begin{center}
\includegraphics[scale=.25]{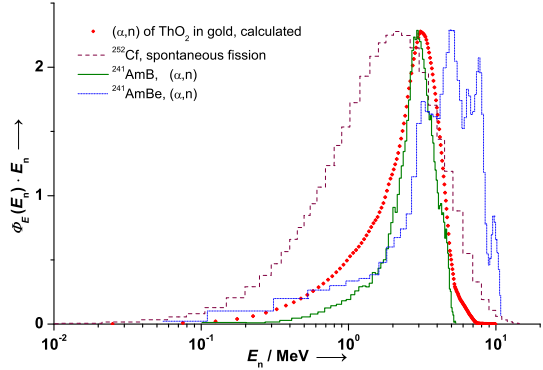}
\caption{Neutron energy spectra of the Borexino $^{228}$Th source and of the three reference sources $^{241}$AmBe, $^{241}$AmB and $^{252}$Cf.}
\label{fig:lethargy-spectra}
\end{center}
\end{figure}

The neutron source strength of the custom-made Borexino $^{228}$Th source was measured at the Physikalisch-Techische Bundesanstalt in Braunschweig, Germany.\\
To measure the neutron emission components of the PTB Bonner sphere spectrometer were used: a spherical proportional counter of type SP9 filled with 200\,kPa $^{3}$He gas and the standard data acquisition system. The SP9 counter made by Centronic Ltd., Croyden, UK, is most sensitive to thermal neutrons ($E$ $<$1\,eV) which produce two charged particles in the reaction $^{3}$He(n,p)$^{3}$H with a Q-value of 764\,keV. Gamma-rays entering the counter also ionize the $^{3}$He gas but with much lower total deposited energy. Therefore the SP9 counter is well suited for n-$\gamma$ discrimination. However, for the Borexino $^{228}$Th source the neutron source strength is expected to be only a very small fraction of the $\gamma$-source strength. In order to investigate the n-$\gamma$ discrimination for such conditions, the SP9 detector was exposed in the PTB irradiation facility to (i) a 11\,GBq $^{137}$Cs source, (ii) a $^{252}$Cf source with neutron source strength of 4$\times$10$^5$ s$^{-1}$ and (iii) the $^{137}$Cs and the $^{252}$Cf sources together to simulate a $\gamma$-n particle ratio of 10$^5$ and higher. It turned out that, compared to the conditions with a lower photon background, the hardware and software thresholds 
had to be increased. Herein, the hardware threshold defines the lowest pulse height fed into the ADC, while the software threshold in the pulse height spectrum in\-di\-cates the separation between $\gamma$-rays and neutrons. Using these settings it could be shown that the large amount of coexistent $\gamma$-radiation does not influence the neutron efficiency of the SP9 counter.\\  
A moderator set-up including a paraffin block with size 50$\times$75$\times$50\,cm$^3$ was used for the determination of the neutron source strength. The SP9 counter and the Borexino $^{228}$Th source were placed inside the moderator block at a distance of 20\,cm, each in an appropriate holder. All measurements performed within this campaign were done with the above described threshold settings.\\
Prior to the $^{228}$Th source measurements at PTB a series of measurements was done with no source in the paraffin block to determine the background. The background events above the threshold have two origins: from external radiation, i.e. events from high energy neutrons from the cosmic radiation which cannot be shielded and from internal radiation (radioactive impurities inside the counter). The latter contribution is only 2\% of the total background \cite{Neu03}. The event rate from neutrons from the cosmic radiation is anti-correlated to the air pressure \cite{Wie02}. However, for these measurements this effect is of minor importance and was not corrected.
\begin{table*}
\begin{center}
	\begin{tabular}{|c|c|cccc|}
	\hline
	reference 				&	spectrum 	&	mean energy	&	source strength		& reference			&	efficiency				\\
	source (rs)				&	references	&	$\langle E \rangle$ [MeV]   &  $B_{\mathrm{rs}}$ [s$^{-1}$]	& date				&	$c_{\mathrm{rs}} = B_{\mathrm{rs}}/\dot{N}_{\mathrm{rs}}$\\	
	\hline	
	$^{241}$AmBe	 		&	\cite{ISO01}&	4.05		&	72829 $\pm$ 1736	&	23.03.2010		&	1806.3 $\pm$ 43.1		\\
	$^{241}$AmB				&	\cite{Mar95}&	2.61		&	149570 $\pm$ 1081	&	08.05.2010		&	1706.4 $\pm$ 12.3		\\
	$^{252}$Cf				&	\cite{ISO01}&	2.13		&	1131906 $\pm$ 16979	&	06.05.2010		&	2063.6 $\pm$ 31.0		\\
	\hline
	\end{tabular}
	\caption{\rm{List of reference sources to determine the spectrum-dependent efficiency of the experimental set-up.}}
	\label{tab:reference-sources}	
\end{center}
\end{table*} 
To determine the spectrum dependent efficiency of the experimental set-up, three different radionuclide sources of known source strength and different neutron spectra, i.e. different mean energy, were used (see Table \ref{tab:reference-sources}). The quantity $c_{\mathrm{rs}}$ is needed as calibration factor, with `rs' for reference source, to calculate the desired source strength $B_{\mathrm{Th}}$ from the measured count rate $\dot{N}_{\mathrm{Th}}$ with the $^{228}$Th source placed in the same position as the reference sources.\\  
MCNP calculations for ThO$_2$ in a gold foil resulted in a neutron energy spectrum with a mean energy of 2.41\,MeV \cite{Tar11}. Neutrons from $\alpha$-particles of the $^{228}$Th decay chain with Fe can be excluded \cite{Lev10}. Only a small neutron contribution from $\alpha$-particles reacting with Zr may be present (see Section \ref{ch:Characteristics of the initial ThCl$_4$ solution}), but they are not taken into account in the calculation. Looking the neutron energy spectra of the reference source in Figure \ref{fig:lethargy-spectra} the mean energy of the $^{241}$AmB source is closest to the mean energy of the $^{228}$Th source. Thus, the above introduced ca\-li\-bra\-tion factor is chosen to be the efficiency value derived with the $^{241}$AmB source, $c_{\mathrm{rs}} = c_{\mathrm{AmB}}$.\\
The measurements to determine the neutron source strength of the custom-made Borexino $^{228}$Th source 
started 4\,d after closing the source and was performed for a total measurement time of $\sim$10\,d.
Thus, equilibrium of all $\alpha$ emitting daughter nuclides from $^{228}$Th can be assumed. Indeed, no increase of the count rate was observed during the entire measurement period.\\
The neutron source strength of the $^{228}$Th source was determined as follows:
  \begin{equation}
  B_{\mathrm{Th}} = f_c \cdot c_{\mathrm{AmB}} \cdot \dot{N}_{\mathrm{Th}}
  \label{eq:source-strength} 
  \end{equation}
with
\begin{subequations}
\begin{align}
f_c &\equiv 1\\
c_{\mathrm{AmB}} &=B_{\mathrm{AmB}}/\dot{N}_{\mathrm{AmB}}\\
\dot{N}_{\mathrm{Th}} &= \dot{N}_{\mathrm{bgTh}} - f_{\mathrm{NTh}} \dot{N}_{\mathrm{bg}}\\
\dot{N}_{\mathrm{AmB}} &= \dot{N}_{\mathrm{bgAmB}} - f_{\mathrm{NAmB}} \dot{N}_{\mathrm{bg}}\\
f_{\mathrm{NTh}} &= f_{\mathrm{NAmB}}\equiv 1
\end{align}
\end{subequations}
Herein $f_c$ is the correction factor to include the uncertainty due to the choice of the factor $c_{\mathrm{rs}}$. The coefficient $c_{\mathrm{AmB}}$ stays for the spectrum-dependent efficiency factor for the $^{241}$AmB source, $\dot{N}_{\mathrm{Th}}$ and $\dot{N}_{\mathrm{AmB}}$ stay for the background corrected count rate of the $^{228}$Th and $^{241}$AmB source. $\dot{N}_{\mathrm{bgTh}}$ and $\dot{N}_{\mathrm{bgAmB}}$ corresponds to the measured count rate with $^{228}$Th and $^{241}$AmB source inserted, including the background of cosmic radiation. $\dot{N}_{\mathrm{bg}}$ is the count rate due to the background events, $f_{\mathrm{NTh}}$ and $f_{\mathrm{NAmB}}$ are the air pressure correction factors.\\
The factors $f_{c}$, $f_{\mathrm{NTh}}$ and $f_{\mathrm{NAmB}}$ are all set to unity but they are needed to calculate a correct uncertainty budget using the Guide to the Expression of Uncertainty in Measurements (GUM) \cite{ISO95}. As it can be seen from Table \ref{tab:reference-sources}, the values for the source dependent efficiency $c_{\mathrm{rs}}$ show a spread of about 20\%. 
Using the maximum difference and a rectangular distribution as estimate for the uncertainty of $f_c$, it follows $f_c=1\pm 0.121$. As was mentioned before, the anti-correlation of background count rate and air pressure was not corrected, but we assign an uncertainty to both correction factors as $f_{NTh} = f_{NAmB} = 1\pm 0.072$. The measured count rate of the background events is $\dot{N}_{\mathrm{bg}} = (7.26 \pm 0.11)$\,h$^{-1}$ and the count rates of the sum of neutrons from the $^{228}$Th source respectively $^{241}$AmB source plus background events are $\dot{N}_{\mathrm{bgTh}} = (20.38 \pm 0.28)$\,h$^{-1}$ and $\dot{N}_{\mathrm{bgAmB}} = (3.1555 \pm 0.0012){\times}10^5$\,h$^{-1}$.\\
By including all measured values and their associated uncertainties the source strength of the custom-made Borexino $^{228}$Th source was determined to be 
\begin{equation*}
  B_{\mathrm{Th}} = (6.59 \pm 0.85)\,\mathrm{s^{-1}}
  \label{eq:source-strength}
\end{equation*}
(reference date: March 1, 2010). The uncertainty assigned to $f_c$ reflects the unsureness in the choice of the calibration factor and it contributes with 87\% to the total uncertainty of $B_{\mathrm{Th}}$ derived with GUM. The next smaller contribution of 9.4\% is connected to the air pressure correction factor $f_{\mathrm{NTh}}$ which is used in the calculation of the net count rate $\dot{N}_{\mathrm{Th}}$  while the thorium source was placed inside the paraffin block. The statistical uncertainty of the count rate $\dot{N}_{\mathrm{bgTh}}$ is only the third largest contribution (2.7\%).

\subsection{Dose rate estimation} 
\label{Dose rate estimation} 

Finally the dose rate of the Borexino 5.41\,MBq $^{228}$Th source was estimated. This is of major importance for a correct handling of the source during calibrations in terms of radiation protection following national law regulations.\\
The expected value of a dose rate $h$ emitted by a source of activity $A$ (unit: GBq) in a distance $r$ (unit: m) is given by
\begin{equation}
  h = k \cdot \frac{A}{r^2}
\label{eq:dose-rate} 
\end{equation}
Herein, the factor $k$ corresponds to a specific radiation constant for a point-like source that is individual for each nuclide. For $^{228}$Th including all its daughter nuclides and assuming secular e\-qui\-li\-brium the factor $k$ is 0.187\,mSv\,m$^2$\,h$^{-1}$\,GBq$^{-1}$ \cite{Ger11}. Then, the expected dose rate in 1\,cm distance from the source becomes $h_{01}=10.1$\,mSv\,h$^{-1}$.\\
The dose rate has been measured by means of dosimeters placed in $\sim$30\,cm distance from the source. The used dosimeters are a Berthold UMo LB123 monitor with two different individually calibrated proportional counter probes of type Berthold LB1236. The instruments were made by Berthold Technologies GmbH \& Co KG, Bad Wildbad, Germany. From the measurements a dose rate of $\sim$12.1\,mSv\,h$^{-1}$ and $\sim$14.3\,mSv\,h$^{-1}$ in 1\,cm distance from the source was deduced. These values are in rather good agreement with the expectation.

\section{Conclusions}

The Borexino $^{228}$Th source described in this paper is unique in terms of its production, characterisation and achieved properties.\\
To our knowledge it is the first time that a custom-made $^{228}$Th source of several MBq activity was successfully produced where a ThCl$_4$ solution was chemically converted into ThO$_2$ on a gold substrate. The need for such a source was given by the fact that radioactive nuclides in commercially available sources are typically embedded in a ceramic matrix (low-Z material) and thus emit neutrons via ($\alpha$,n) reactions. These neutrons can induce unwanted background signals in the own or neighboring experiments looking for rare events. By using materials with a higher ($\alpha$,n) energy threshold such as gold their neutron source strength can be suppressed. Other possibilities to produce a $^{228}$Th source with a reduced neutron source strength were not explored but might be also complex. First, one could try to produce an inter-metallic alloy such as Th$_2$Gd \cite{Bau58} by arc melting under reduced argon atmosphere. Second, if exclusively 2.614\,MeV $\gamma$-rays are of interest a non-commercial $^{208}$Bi could be produced. $^{208}$Bi is an artificial unstable $\beta^+$ emitter ($\tau$=3.65$\times$10$^5$\,y) that emits purely 2.614\,MeV and 0.511\,MeV photons, the last ones from positron annihilation. Thus, there are no ($\alpha$,n) reactions and neutron emissions at all. However, $^{208}$Bi could only be produced via fast neutron activation (energy threshold: 7.5\,MeV) at a reactor site.\\ 
The Borexino $^{228}$Th source was fully characterized by means of its $\gamma$-activity, neutron source strength, dose rate and sealing properties. The activity $A$=(5.41$\pm$0.30)\,MBq of the $^{228}$Th source was measured by comparison with a calibrated $^{228}$Th source. The neutron source strength $B_{\mathrm{Th}}$ was measured to be (6.59$\pm$0.85)\,s$^{-1}$. ($\alpha$,n) reactions due to the presence of $^{17}$O, $^{18}$O, Zr and other impurities within the used gold encapsulation might have led to this small but still present amount of escaping neutrons. However, the ratio $B_{\mathrm{Th}}/A$=(1.22$\pm$0.17)$\times$10$^{-3}$\,s$^{-1}$\,kBq$^{-1}$ is several times less than from commercial sources based on ceramic compositions according to measurements and MCNP-based calculations using the code SOURCES 4A \cite{Tar11}. The obtained neutron suppression is sufficient when comparing the obtained value for $B_{\mathrm{Th}}$ with the neutron background from the walls at LNGS: The measured neutron flux in hall A is ($3.7\pm 0.2)\times $10$^{-2}$\,s$^{-1}$ m$^{-2}$ \cite{Bel89}. With a hall size of about 100$\times$20$\times$18\,m$^3$ one obtains an integrated flux that is larger than the neutron source strength of the Borexino $^{228}$Th source. Moreover, taking into account the ratio 3$\times$10$^5$:1 of 2.614\,MeV $\gamma$-rays to emitted number of neutrons, the determination of the neutron source strength set a milestone in terms of background suppression for $^3$He proportional gas counters.\\ 
So far, the here presented $^{228}$Th source has been used for the external calibration of the Borexino detector. Due to the precise characterisation and the relatively long life time of $^{228}$Th ($\tau$=2.76\,y) the Borexino $^{228}$Th source might also be useful for other rare-event experiments in underground laboratories that have to cope with an intrinsic external background.

\section{Acknowledgements}
  \label{ch:Acknowledgements}

The Max Planck Institut f\"ur Kernphysik sincerely thanks the Paul Scherrer Institut in Villigen, Switzerland, the Institut f\"ur Kernchemie in Mainz, Germany, and the Physikalisch-Technische Bundesanstalt in Braunschweig, Germany, for the scientific cooperation. The authors thank L. Lembo, A. Giampaoli, M. Laubenstein and D.~Budj\'a\v{s} for their support and contribution. W. Maneschg thanks the company Eckert und Ziegler Nuclitec GmbH for its constant courtesy. This work was supported by INTAS (Project Nr. 05-1000008-7996) and by DFG within the SFB Transregio 27 "Neutrinos and Beyond".



\end{document}